\documentclass[preprint]{aastex}

\shorttitle{The BOES spectropolarimeter for Zeeman measurements of
stellar magnetic fields}
\shortauthors{Kim.~K.-M. et al.}

\begin{document}

\title{The BOES spectropolarimeter for Zeeman measurements of
stellar magnetic fields}

\author{Kang-Min Kim\altaffilmark{1},
Inwoo Han\altaffilmark{1}, Gennady G. Valyavin\altaffilmark{1},
Sergei Plachinda\altaffilmark{2}, Jeong Gyun Jang\altaffilmark{1},
Be-Ho Jang\altaffilmark{1}, Hyeon Cheol Seong\altaffilmark{1},
Byeong-Cheol Lee\altaffilmark{1,3}, Dong-Il
Kang\altaffilmark{1,4}, Byeong-Gon Park\altaffilmark{1}, Tae Seog
Yoon\altaffilmark{3}, Steven S. Vogt\altaffilmark{5}}

\altaffiltext{1}{Korea Astronomy and Space Science Institute,
61-1, Whaam-Dong, Youseong-Gu, Daejeon 305-348, Korea}

\altaffiltext{2}{Crimean Astrophysical Observatory, Nauchny,
Crimea, 98409, Ukraine}

\altaffiltext{3}{Department of Astronomy and Atmospheric Sciences,
Kyungpook National Univ., Daegu 702-701, Korea}

\altaffiltext{4}{Department of Earth Science Education, Korea
National Univ. of Education, Chung-Buk, 363-791, Korea}

\altaffiltext{5}{University of California Observatories/Lick
Observatory, University of California at Santa Cruz, Santa Cruz,
CA 95064, USA}
%

\begin{abstract}
We introduce a new polarimeter installed on the high-resolution
fiber-fed echelle spectrograph (called BOES) of the 1.8-m
telescope at the Bohyunsan Optical Astronomy Observatory,
Korea. The instrument is intended to measure stellar magnetic
fields with high-resolution (R $\sim$ 60000) spectropolarimetric
observations of intrinsic polarization in spectral lines. In this
paper we describe the spectropolarimeter and present
test observations of the longitudinal magnetic fields in some well-studied
F-B main sequence magnetic stars ($m_v < 8.8^m$).  The results
demonstrate that the instrument has a high precision ability of
detecting the fields of these stars with typical accuracies
ranged from about 2 to a few tens of gauss.

\end{abstract}

\section{Introduction}
\vspace*{0.5cm}

The presence of intrinsic linear and circular polarizations in
spectra of stellar objects provides an important information for diagnostics
of their magnetism, wind surroundings, atmospheric inhomogeneities and other
properties. For example, non-zero continuum linear polarization due to
Thomson and Rayleigh scattering demonstrates the presence of
non-symmetric patterns in the distribution of atmospheric or wind
medium. The broad-band circular polarization as well as circular
and linear polarizations in spectral lines exhibit information on
the magnetic fields.  The spectropolarimetric observation is therefore one
of the most important tools for the experimental studies of stellar
magnetism. To argue this statement, we start our consideration
with a brief presentation of the most important historical results
that have come to us from stellar spectropolarimetry.

\begin{itemize}
\item
{\it Magnetism of chemically peculiar main sequence stars.} Strong
magnetic fields up to several tens of kilogauss have been detected
and studied on a large sample of chemically peculiar (CP) stars.
In contrast to non-regular, localized magnetic fields of
solar-type stars, statistical properties of the fields in CP stars
and results of detailed modeling of field geometries in individual
objects are generally consistent with the picture of smooth,
roughly dipolar magnetic field, inclined with respect to the
stellar axis of rotation \citep{LA01}.

\item
{\it Magnetism of early-type pulsating and hot O-stars.} Recently,
comparatively weak regular magnetic fields have been found also in
hot, massive $\beta$ Cep-type and some other pulsating stars
\citep{HBS06}, and O-type stars \citep{WFD06}. Magnetic fields of
these stars have also several morphological differences from
magnetic fields of the Sun and other late-type stars.

\item
{\it Magnetism of late-type, convective stars.} Practically all
manifestations of solar activity (chromosphere and corona, plages
and spots, flares, etc.) are related to magnetic fields and their
interaction with differential rotation and convection.
Spectropolarimetric studies of fragmented and global magnetic
fields make it possible to extend our knowledge about these
processes to another solar-type and cooler convective stars.

At the moment, the presence of global magnetic fields from a few
to some tens of gauss has been established in convective stars of
F9--M3 spectral types among I--V luminosity classes
\citep{pla99,pla00,don03,pla04a,pla04b,pet05}.
\end{itemize}

This brief and, of course, not complete presentation illustrates
the importance of spectropolarimetry as effective
observational tool for studying the stellar magnetism.

Nowadays there is a large collection of portable \citep[for
instance]{EMD98} or stationary spectropolarimeters installed at
different telescope.  The 2--3\,m class telescopes equipped
with stationary spectropolarimeters are:\\

\noindent
2.0m telescope, Pic du Midi, France \\
2.6m telescope, Crimea, Ukraine \\
2.6m Nordic telescope, La Palma, Spain \\
2.6m telescope, McDonald, USA \\
3.0m Italy national telescope, Italy \\
3.6m Canada-France-Hawaii telescope \\

Among the most known versions of spectropolarimeters installed at
large telescopes are the imager/spectrograph/polarimeter FORS1
\citep{AFF98} of the ESO, VLT; spectropolarimeters at the AAT
\citep{B89} and Keck \citep{GCP95}. Very recently a low-resolution
polarimeter was also installed at the 6-m Russian telescope
\citep{NVF02}. New magnetic weak-field main sequence and
degenerate stars were detected/studied with these instruments
(see, for instance, a review by \citet{P99} also
\citet{AZC04,VBF06,Wad03} and references therein).

Recently designed high-resolution fiber-fed spectropolarimeters
with intermediate class telescopes such as the MUSICOS
\citep{DCW99}, or ESPaDOnS \citep{MD03} have demonstrated the best
characteristics in obtaining the polarization spectra and
measuring the stellar magnetic fields that created new generation
of stellar spectropolarimeters. Typical accuracies of the field
measurements with these instruments range from about 1 to a few
tens of gauss that corresponds to the best instrumental level ever
achieved with polarimeters of the 2--3m class telescopes
\citep[and references therein]{WDL00}.

In this paper we introduce a new fiber-fed spectropolarimeter
installed on the 1.8-m telescope of the Bohyunsan Optical
Astronomy Observatory (BOAO) in Korea.  This polarimeter provides
high-resolution (R $\sim$ 60000) observations of all four Stokes
Parameters and intended for high-precision Zeeman measurements of
stellar magnetic fields with typical accuracies from about 1 to a
few tens of gauss depending on the spectral characteristics of
studied stars. Here we present the structure of this polarimeter
and results of the longitudinal magnetic fields measurements for
some well-studied magnetic stars.  Observation of linear
polarization is also available. However, due to poor weather
conditions at the BOAO site and the presence of significant light
pollution, observations of linear polarization are complicated and
not so quite effective.  Now, we are focused mainly on
observations of circular polarization with this spectropolaimeter.

\section{The BOES spectropolarimeter: basic principles and design. }

The BOES (\textbf{B}oa\textbf{O} \textbf{E}chelle
\textbf{S}pectrograph) is a high throughput, versatile, fiber-fed
prism cross-dispersed echelle spectrograph installed at the 1.8-m
telescope of the BOAO. Using a 2k x 4k CCD, the BOES can obtain a
spectra simultaneously over a wide wavelength range of 3500 --
10500 $\AA$ with a throughput (resolution x slit width in arcsec.)
around 125,000. It has nine fibers with core diameter from 80 to
300 $\mu$m, corresponding field of views and resolutions
($\lambda$/$\Delta \lambda$) are 1.1 $\sim$ 4.3 arcsec. and 30000
$\sim$ 90000 respectively.

\begin{figure}
\centering
\includegraphics[width=8.5cm, height=10cm, angle=0]{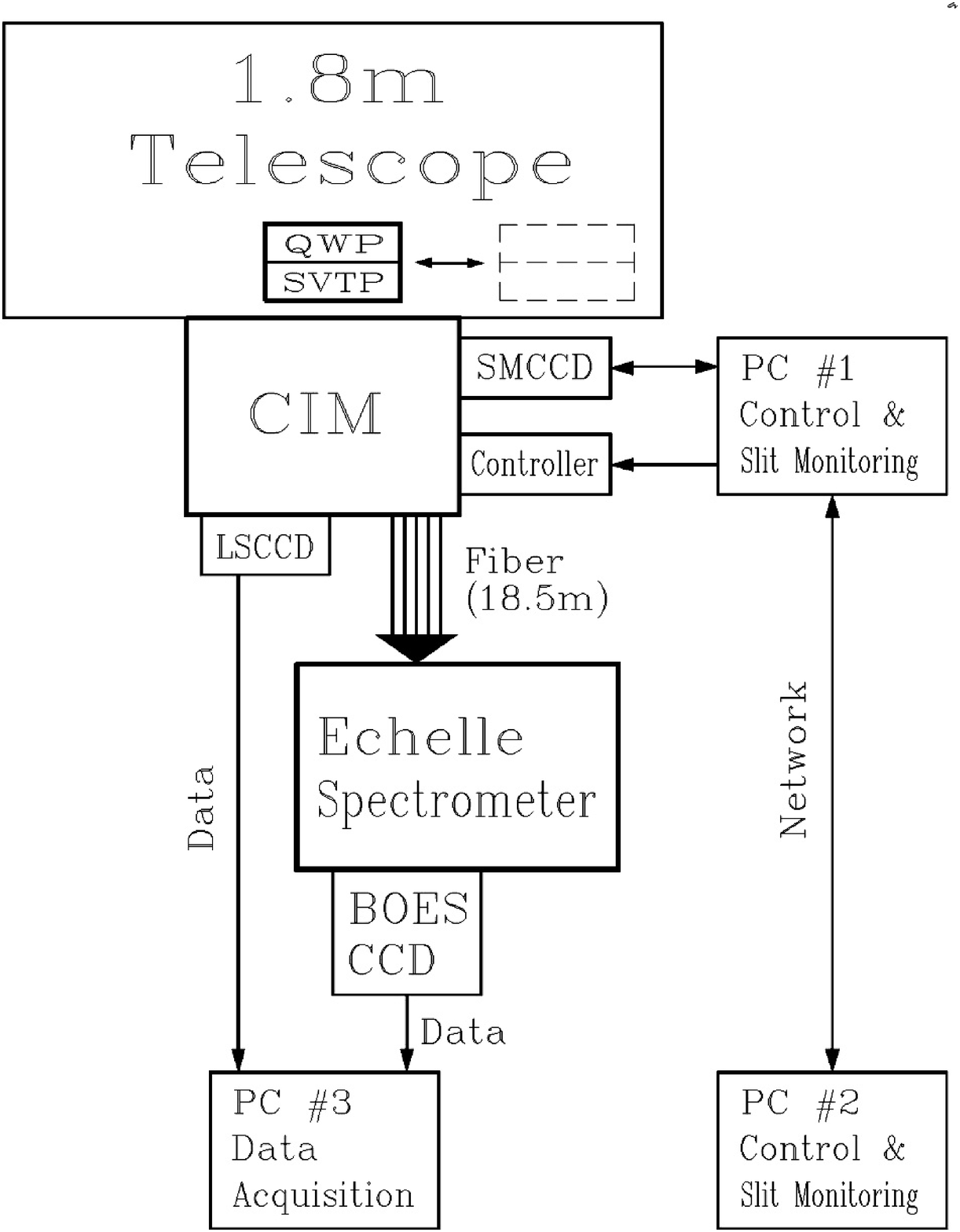}
\vspace*{0.5cm} \caption{ The block diagram of the BOES.}
\label{fig1}
\end{figure}

\begin{figure}
\centering
\includegraphics[width=8.5cm, height=7.5cm, angle=0]{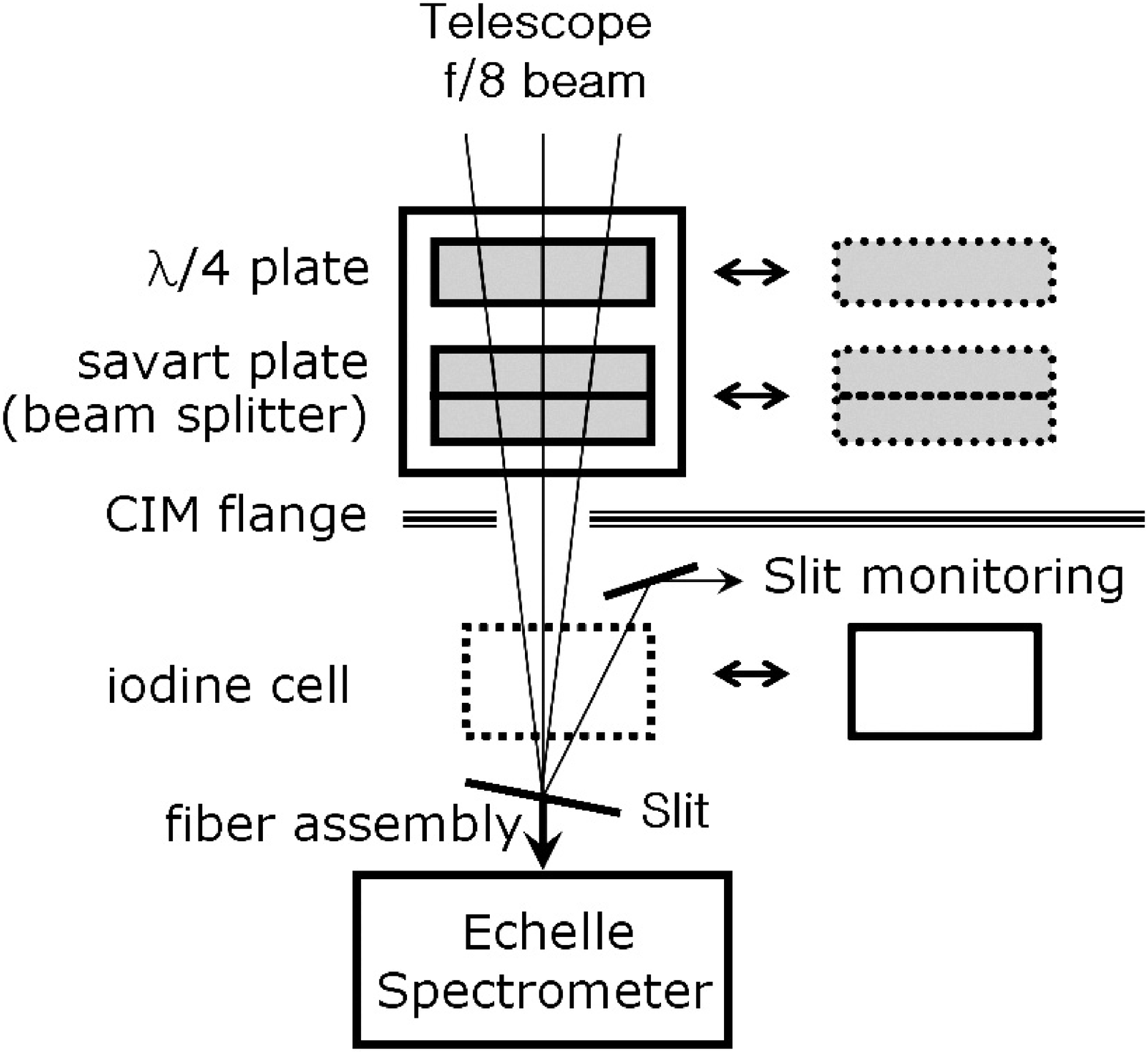}
\vspace*{0.5cm} \caption{
 The optical layout of the polarimetric
optics in front of the CIM of the 1.8-m telescope } \label{fig2}
\end{figure}

\begin{figure}
\centering
\includegraphics[width=8.5cm, height=11cm, angle=0]{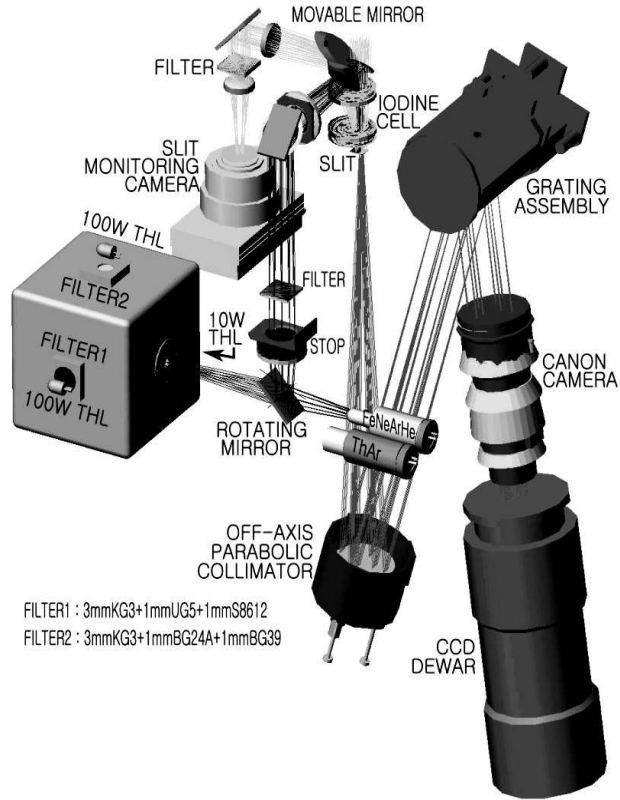}
\vspace*{0.5cm} \caption{ The optical layout of the BOES CIM. The
long slit spectrograph, slit monitoring device and the calibration
lamp unit can be seen at the right, left and front part
respectively in this figure.  Fiber input of the fiber assembly
(omitted in this figure) is located at the slit position (see the
Fig.2).
 } \label{fig3}
\end{figure}

\begin{figure}
\centering
\includegraphics[width=8.5cm, height=5cm, angle=0]{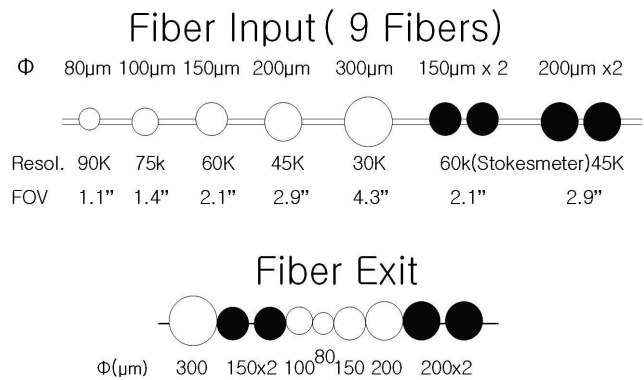}
\vspace*{0.5cm} \caption{ The scheme of the fiber input and exit
in the BOES. Resolutions and fields of view are presented.  The
black circles denote the pairs of fiber for the
spectropolarimeter.} \label{fig4}
\end{figure}

\begin{figure}
\centering
\includegraphics[width=8.5cm, height=7.5cm, angle=0]{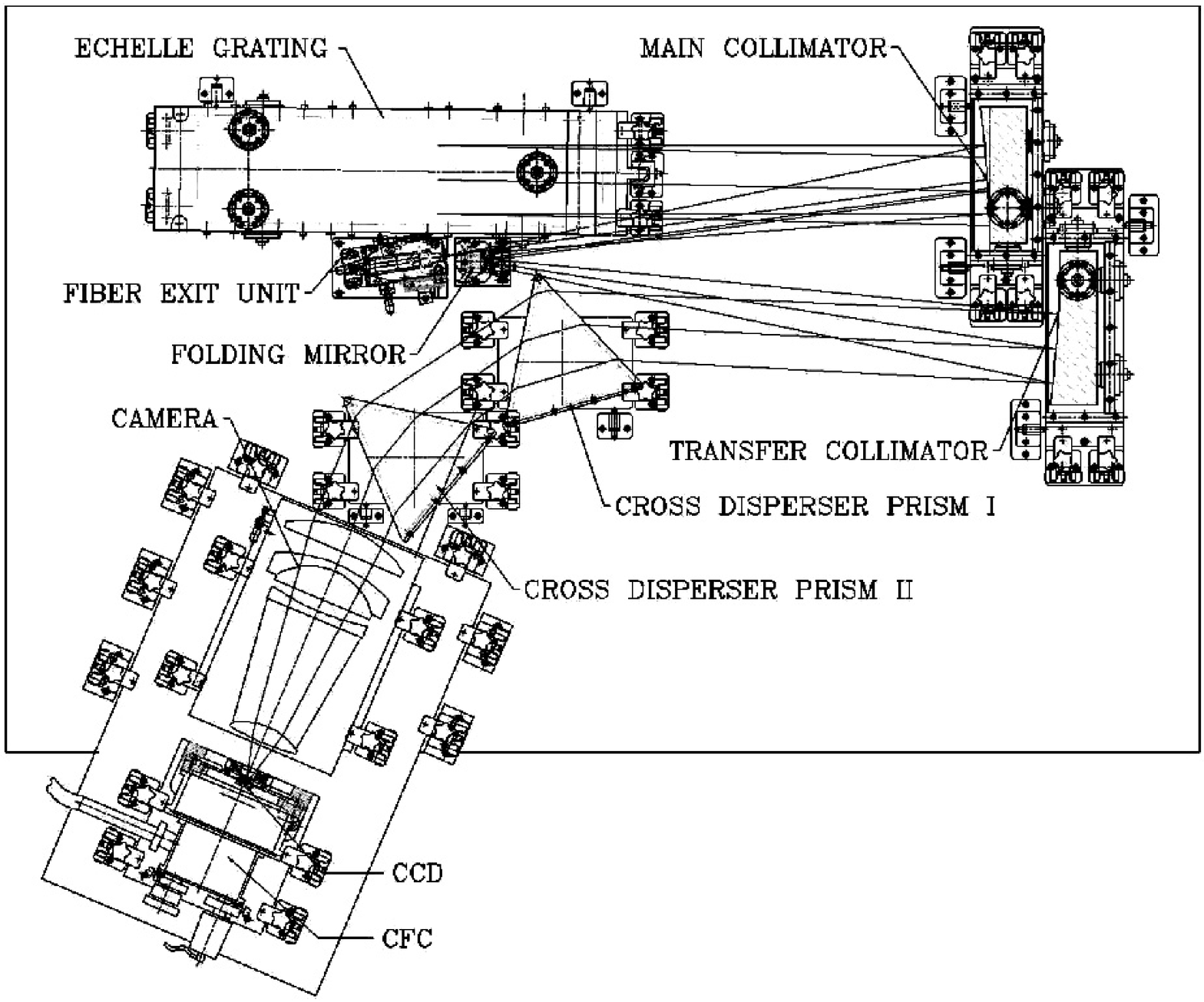}
\vspace*{0.5cm} \caption{ The spectrometer part of the BOES on the
optical bench.   The black anodized enclosure to block the dust
and the turbulence from air circulation, and a baffle screen to
block the stray light were omitted in this figure.  CFC in this
figure denotes Continuous Flow Cryostat dewar.} \label{fig5}
\end{figure}

The block diagram of the BOES is presented in Fig.\,\ref{fig1}.
The instrument consists of three main parts; 1) polarization
optics (BOESP, Fig.\,\ref{fig2}) containing a rotatable quarter
wave plate(QWP) and a Savart plate (SVTP) as a beam splitter, 2)
CIM (Cassegrain Interface Module, Fig.\,\ref{fig3}) with a light
transmitting fiber set (Fig.\,\ref{fig4}), and 3) a spectrometer
(i.e. a bench mounted echelle spectrograph, Fig.\,\ref{fig5}). The
18.5 m length optical fibers transmit the star light accumulated
at the Cassegrain focus to the spectrometer room where the
temperature and humidity maintained to 20 $\pm$ 0.5 $^\circ$C and
below 50 $\%$ respectively.

In addition to the high resolution echelle mode, BOES is equipped
with a medium dispersion long slit spectrograph (called LS)  in
the CIM (Fig.\,\ref{fig3}).  The LS can obtain a spectrum of a
linear reciprocal dispersion of 19 to 217 $\AA$/mm (0.45 to 5.2
$\AA$/px) with 3.6 arcmin. slit length. The device switching
between BOES (ordinary spectroscopic mode), BOESP
(spectropolarimetric mode) and LS (long slit spectroscopic mode)
can be done within 10 seconds.

As shown in Fig.\,\ref{fig1}, the BOES uses three CCDs : a slit
monitoring CCD (called SMCCD), LS CCD and the BOES CCD.  For the
SMCCD, we use the Quantix 57 camera which has EEV 57-10 CCD chip
(530 x 526, 13 $\mu$m pixel) with 2.7 arcmin.  field of view.  The
LS CCD has Tek 1024 chip (1024 x 1024, 24 $\mu$ pixel) with the
readout noise of 7.4 e$^-$. The BOES CCD is a grade zero class E2V
CCD44-82 chip (2048 x 4096, 15 $\mu$m pixel) with  4.0 e$^-$
readout noise.

Three personal computers (PC) are used to control the instrument.
The PC $\sharp$1 connected to the CIM through RS232C controls the
CIM. The observer can monitor the slit image and control the CIM
by the PC $\sharp$2 connected to the PC $\sharp$1 by network. And
the data acquisition from the BOES CCD or LS CCD is performed by a
Linux based PC $\sharp$3.

\subsection{Polarimetric optics}
The polarimetric analyzer that we use in the spectropolarimetric
mode consists of a rotatable QWP for polarimetric modulation, and
Savart plate as a beam-splitter. In the present design we have
decided to use the polymer QWP \citep{sam04} instead frequently
used Quartz/MgF$_{2}$ crystal wave plate or Fresnel rhombs. It is
well-known fact, that the Quartz/MgF$_{2}$ QWP gives significant
artificial polarization ripples of 0.05 to 2 $\%$ (\cite{HH96},
\cite{DCW99}) due to the interference within the cemented layers
of the retarder. And well known solution based on Fresnel rhombs
can not easily be used in our design due to the considerable size
of these optics. In the same time, the used polimer QWP gives the
wave retard of 0.25 $\pm$ 0.007$\lambda$ in 4000 $\sim$ 8000 $\AA$
range \citep{sam04}, and the ripple below 0.1 $\%$ \citep{ike03}.
Such good parameters of the plate make it possible to use it
alternatively. With these optical elements the available working
wavelengths of the polarimeter is 4000 -- 8000 $\AA$.

The optics are installed at the Cassegrian focus in front of the fiber
input as shown in Fig.\,\ref{fig2}. It is mounted on rotary stages
to be removable from the optical axis in case of ordinary spectral
observations. The spectropolarimetric observations are available
with resolving power of 45000 or 60000.

\begin{figure}
\centering
\includegraphics[width=8.5cm, height=8.5cm, angle=0]{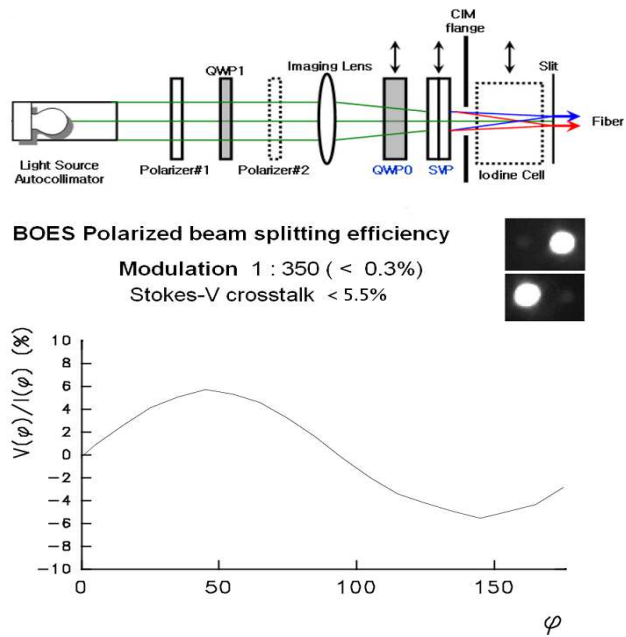}
\vspace*{0.5cm} \caption{ The optical layout of the modulation and
crosstalk measurements (the upper plot), and the results of the
crosstalk measurements (the lower plot).  While measuring the
crosstalk, the QWP\#1 was removed. } \label{fig6}
\end{figure}

Before each run of the polarimetric observations, the adjustment
of the polarimeter is examined using a laboratory source of the
light (for example, similar procedure is described in details by
\cite{plach05}). Some important examples of these tests are
presented in Fig.\,\ref{fig6}, where the upper plot illustrates
the analyzer assembled for measurements of the modulation by the
QWP in observations of circular polarization. Artificial
circularly polarized light is created by the combination of
additional quarter-wave plate (QWP1) and polarizer (polarizer\#1)
as shown in the figure. The efficiency of the modulation is better
than 99.7\% . The lower plot (graph) illustrates the results of
crosstalk measurements obtained at different orientation angles
(the horizontal axis) of polarizer\#1  (the QWP1 is removed in
this case). As one can see, maximum crosstalk (vertical axis in
the figure) which characterizes artificial circular polarization
from linearly polarized light is not higher than 5.5\% . To our
knowledge, this result is practically standard for modern
polarimeters.

\subsection{The CIM and the fiber assembly}

As the design concept of the CIM and the fiber assembly was
described in \citet{kea02}, we do not mention it here in details.
Since then, we revised the flat-fielding lamps and the fiber
assembly that was adopted for polarimetric observations.  For the
white balance of the flat fielding, we use three tungsten halogen
lamps with an integration sphere -- a 10 watts lamp with a half
inch diameter aperture for the red wavelengths, a 100 watts with 1
inch diameter filters (3mmKG3 + 1mmBG24A + 1mmBG39) for the medium
wavelengths, and a 100 watts with 2 inch diameter filters (3mmKG3
+ 1mmUG5 + 1mmS8612) for the blue region (see the cube type
integrating sphere in Fig.\,\ref{fig3}).

The BOES is furnished with nine fibers of 80, 100, 150, 200 and
300 $\mu$m in core diameters that provides corresponding spectral
resolutions $\lambda$/$\Delta \lambda \sim$\,\,90000, 75000,
60000, 45000 and 30000 respectively (Fig.\,\ref{fig4}). Five of
them are used for ordinary spectroscopy, and two pairs of 150 and
200 $\mu$m fibers for spectropolarimetry. At the fiber input,
fibers in each pair are separated by 500 $\mu$m distance that
corresponds to the separation of the beams split by the
polarimetric analyzer. At the fiber exit, these are separated to
380 $\mu$m to avoid overlapping of spectral orders. All the fibers
except the 80 $\mu$m one (STU, \citet{sch98}) are FBP
\citep{pol04}, transmission of which is improved relative to STU
fiber, especially in the blue region.

\subsection{Spectrometer part}

The spectrometer (BOES) was designed by SSV and is shown in Figure
\,\ref{fig5}. It is a quasi-Littrow configuration in the
dual-white-pupil (DWP) configuration pioneered by UVES
\citep{dek92} on the VLT. Other similar style DWP spectrometers
are FOCES \citep{pfe98} of the 2.2-m telescope at Calar Alto
Observatory, HRS \citep{tul98} on the HET, and FEROS \citep{kau98}
at ESO.

In the BOES design, the f/8 beam exiting the fiber is collimated
by the main off-axis collimator into a 136 mm diameter beam which
is then dispersed by a 41.59 gr/mm R4 echelle of size 203 x 813
mm. This echelle is a replica of the master ruled for the blue
side of UVES. The dispersed light from the echelle returns in
quasi-Littrow mode (0.6$^\circ$ out of plane) back to the main
collimator and, via the folding mirror, to the transfer
collimator. The transfer collimator is identical to the main
collimator and the optical axes of both are co-linear. The
transfer collimator corrects, to a high degree, the aberrations
introduced by the main collimator and provides a white pupil near
the cross-disperser to minimize the required clear apertures of
the prisms and camera. Both collimators are cut from a common
f/1.8 parent parabola of 600 mm diameter. All the mirrors have
durable high-reflectance silver coatings.

Most if not all previous versions of DWP spectrometers
incorporated a toroidal rear surface in the optical train (near
the focal plane) to counteract astigmatism introduced by the
quasi-Littrow white pupil collimation combination. However, we
found that slightly pistoning the transfer collimator provides an
equally viable solution that eliminates the need for this aspheric
optic.

For the cross disperser, we adopted the use of a pair of
55$^\circ$ prisms instead of a grating. Prisms provide more
uniform order separation and higher efficiency across the very
wide spectral bandpass of BOES. We used S-BSL7Y, a BK-7-like glass
with enhanced ultraviolet transmission and available from Ohara,
Inc. It was selected on the basis of its unusually high ratio of
red to blue dispersion, creating a more uniform order separation
across the echelle format, and thus more efficient order packing.
The f/1.6 camera has an effective focal length of 389 mm and
consists of six spherical lenses in 3 groups. It works over about
a 9.5$^\circ$ diameter field of view and spans the entire 3500 --
10,500 $\AA$ range with adequate image quality. All prisms and
camera lenses were treated with wide-band anti-reflection
coatings, which provide reflectance below 1.5$\%$ across the 3500
to 10,500 $\AA$ range. Overall, 86 spectral orders, from the 46th
to the 131st, are captured on the CCD. Inter-order stray light
appears to be less than 2$\%$ of the intensities of the
neighboring orders.

\subsection{Efficiency and stability of the BOES}

\begin{figure}
\centering
\includegraphics[width=8.5cm, height=7.0cm, angle=0]{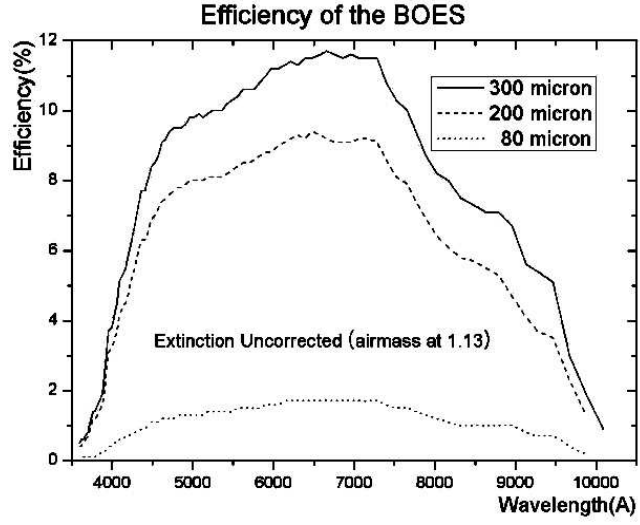}
\vspace*{0.5cm} \caption{ Measured efficiency of the (BOES +
telescope + air extinction) for different single fibers without
the correction of the light cutoff at the fiber input while the
star at 1.13 airmass with the seeing around 2.3 arcseconds on 3rd
Nov., 2003. } \label{fig7}
\end{figure}

Fig.\,\ref{fig7} illustrates the typical efficiency of the BOES
measured on 3rd November 2003 with different (300, 200 and 80
$\mu$m in diameter) single fibers. This shows maximum efficiency
is up to 12$\%$ including the light loss due to the atmospheric
extinction, the reflectance of the telescope and the light cutoff
at the fiber input. While measuring the efficiency, the target
star was at the airmass of 1.13 with the seeing size around 2.3
arcseconds. The small efficiency of the 80 $\mu$m fiber comes from
the light cutoff at the fiber input due to the small diameter
\citep{kea02}.

\begin{figure}
\centering
\includegraphics[width=8.5cm, height=6.0cm, angle=0]{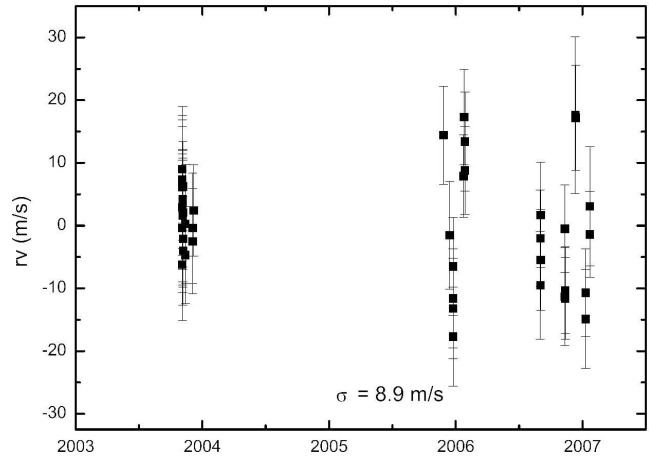}
\vspace*{0.9cm} \caption{ Radial velocity variation in Tau Ceti
measured by the BOES for three years. } \label{fig8}
\end{figure}

Good mechanical stability of the spectrograph makes it possible to
use the instrument in a wide range of observational programs, for
example, asteroseismology as well as Zeeman observations.
Fig.\,\ref{fig8} illustrates the typical accuracy of the stellar
radial velocity measurements achieved in observations with the
BOES equipped with an iodine cell.  The standard star Tau Ceti
(G8V) exhibits no any variations in radial velocities within
typical accuracy about 9 m/s on three years time base.

\section{Measurements of stellar magnetic fields with the BOES}

\subsection{Preliminary remarks}

Directly, stellar magnetic fields are detected mainly through the
observations of the Zeeman effect in spectral lines. According to
basic physical principles (for more details see, for example,
\citet{LA80}) if an atom is placed in a magnetic field $B$, its
individual energy levels are split into 2J\,+\,1 sub-levels
separated by energy $\Delta E = ge \hslash B / 2mc$, where $g$ is
the Lande factor. As a result, stellar magneto-sensitive spectral
lines are split into a number of $\pi-$ and $\sigma-$ components
which are polarized depending on the orientation of the magnetic
field relative to the observer.

In a longitudinal (parallel to the line of sight) magnetic field
the $\sigma-$ components, which are generally displaced
symmetrically to shorter and longer wavelengths relative to the
non-shifted central $\pi-$ components of spectral lines, have
opposite circular polarizations. Thus by observing circular
polarization in the spectral lines we are able to measure the
longitudinal component of stellar magnetic field. To reconstruct
full vector of the field, additional observations of linearly
polarized $\pi-$ and $\sigma-$ components are needed. These
observations allow to estimate a transverse field component that
together with observations of the longitudinal component give full
vector magnetic field averaged over the stellar disc.
In this paper, we discuss observations of the longitudinal fields only.

The averaged circularly polarized $\sigma-$ components are displaced relative
to the rest wavelength $\lambda_0$ of a spectral line by a factor of

\begin{equation}
\Delta\lambda_B = \pm 4.67 \times 10^{-13}z\lambda^2B_l,
\end{equation}

\noindent where $B_l$ is the longitudinal field component in
gauss, $z$ is the effective Lande factor, $\lambda$ is the
wavelength in $\AA$. Due to the fact that these displaced
components have opposite circular polarizations and are shifted
relative to each other, in the stellar circular polarization
spectra (Stokes-V spectra, or V spectra) they form non-zero
features within magneto-sensitive spectral lines. In most cases of
spectropolarimetric observations these features show well-known
S-shaped features (Fig.\,\ref{fig9}) at the cores of spectral
lines. Their amplitudes and forms depend on magnetic field
strengths, Lande factors, gradients of the line profiles and
spectral resolution of a spectrograph:

\begin{equation}
V \sim z B_l (dI/d\lambda)
\end{equation}

\noindent where $dI/d\lambda $ describes the gradient of a
spectral line convolved with instrumental function of a
spectrograph \citep[for instance]{LA01}. The mean longitudinal
magnetic field can be estimated by applying model methods of
Doppler-Zeeman spectropolarimetric tomography to the analysis of
the Stokes-V spectra \citep[for instance]{EJB02}, or the LSD
method \citep{DSC97}. Alternatively (and traditionally)
longitudinal magnetic fields can simply be measured via analysis
of the displacement (1) between the positions of spectral lines in
the spectra of opposite circular polarizations split by analyzer.

\subsection{Obtaining circular polarization spectra: observations
and data reduction}

Each exposure in inobservations of circular polarization with the
BOES yields two spectra on the CCD -- one from the ordinary beam
and the other from the extraordinary beam split by the analyzer.
In case of an ideal spectropolarimeter (which has no any intrinsic
distortion factors), this single exposure obtained at one of the
orthogonal orientations of the quarter-wave plate would
practically be enough to build V-Stokes spectra and to measure the
magnetic field. However, due to the presence of a large number of
instrumental biases such as pixel-to-pixel inhomogeneity, slightly
different dispersion relationships at each of the split beams and
other factors, it is recommended to obtain an additional exposure
with inverted sign of the polarization effects by rotating the
quarter-wave plate by $90^\circ$. Such a technique makes it
possible to reconstruct the Stokes-V spectra in pixel coordinate
system separately for the spectra of the ordinary and
extraordinary beams that significantly increases the quality of
the output results (for details see, for instance, \citet{pla99},
\citet{BSWLM02} or \citet{AZC04}).

Due to the above mentioned reasons and in order to increase
reliability of our observations of circular polarization, {\it one
observation} with the BOES consists of four short, consecutive
exposures at two orthogonal orientations of the quarter-wave plate
(the sequence of its position angles is $+45^\circ , -45^\circ ,
-45^\circ , +45^\circ$). Assuming \textit{a priori} that the
time-scale of possible physical variability of polarization
features in spectra of non-degenerate stars would be as short as a
few hours, we usually set the integration time for each of the
individual exposures from a few minutes to 0.5 hours depending on
stellar magnitude and sky conditions.

The data reduction was processed with IRAF. The procedures are
mostly standard including the following steps: cosmic ray hits
removal, electronic bias substraction, flat-fielding, 2-D
wavelength calibration, sky background substraction, spectrum
extraction. As an output result we obtained a series of pairs of
left and right circular polarized spectra that we used to build
V-Stokes spectra and measure stellar longitudinal magnetic fields.
We obtained the individual $Stokes-V$ spectra for each of the
echelle spectral orders by applying the technique presented by
\citet{BSWLM02}.

\begin{figure}
\centering
\includegraphics[width=8.5cm, height=7.5cm, angle=0]{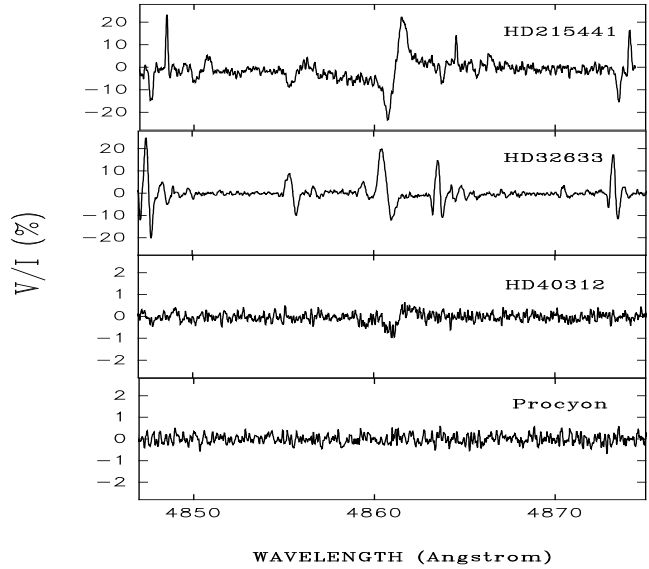}
\vspace*{0.5cm} \caption{Stokes $V$ spectra (from top to bottom) of
the stars HD215441, HD32633, HD40312 and the star HD61421 (Procyon) in
at the H$\beta$ line region.} \label{fig9}
\end{figure}

\begin{figure}
\centering
\includegraphics[width=8.5cm, height=7.5cm, angle=0]{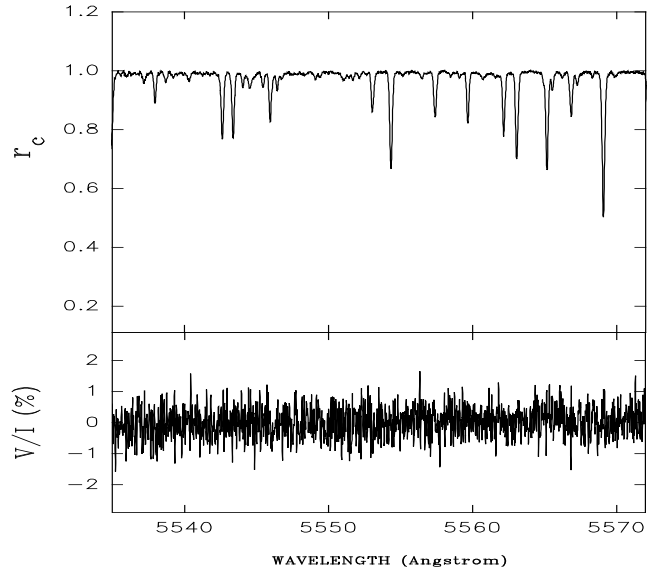}
\vspace*{0.5cm} \caption{Another fragment of Stokes $I$ and $V$
spectra of Procyon around $\lambda 5500$.
There are no any signatures of artificial circular polarization
features in all the studied wavelength region. } \label{fig10}
\end{figure}

In order to illustrate results of the reduction in our first
polarimetric observations with the BOES, we present fragments of
circular polarization spectra of the magnetic stars HD215441,
HD32633, HD40312 and the star HD61421 (Procyon) in
Figs.\,\ref{fig9},\ref{fig10}. We have chosen these stars as the
remarkable cases of typical magnetic stars (HD215441, HD32633,
HD40312) having polarization features of different intensities
(details on these stars are also presented below) and zero-field
(Procyon) star. As one can see (Fig.\,\ref{fig9}), the circular
polarization of different intensities can easily be resolved and
studied with our polarimeter.

Examination of the zero-field star Procyon has revealed no any
artificial circular polarization features at a characteristic level of
about 0.5\% within the working wavelength region from 4000\AA\, to
8000\AA. The lower plot in Fig.\,\ref{fig9} demonstrates zero polarization of
Procyon at the H$\beta$ region. An example of another wavelength region is
presented in Fig.\,\ref{fig10}.

\subsection{Linear polarization}

Finally, before presentation of the results of the longitudinal
magnetic field measurements in the standard stars, we would also
like briefly discuss the possibility to obtain $Stokes-Q/U$
(linear polarization) spectra that is necessary for measurements
of the transverse magnetic fields. It should be noted, that linear
polarization features in spectral lines due to Zeeman effect are
intrinsically weaker than the corresponding circular polarization
features. This strongly complicates observations of linear
polarization with 2-m class telescopes. To simplify the solution,
special methods of data reduction \citep{DSC97} and the LSD method
for obtaining averaged per all spectral lines linear polarization
\citep{DSC97} should be applied. In this paper we do not discuss
these details addressing them for further special consideration of
the problem in application to the BOES. Here we just briefly
present and illustrate test observations of the $Stokes-Q/U$
spectra with the BOES.

There are several designs of polarimetric optics for obtaining all
$Stokes$ parameters including $Stokes-Q/U$. Due to technical
limitations, for linear polarization measurements we have chosen
simplest configuration of polarimetric optics similar to that
presented by \citet{NVF02}. According to their scheme, the
$Stokes-Q/U$ spectra can simply be obtained by using the beam
splitter only (without the QWP). In this case, the necessary
observational basis is achieved by rotation of the cassegrain
assembly with obtaining of four consecutive exposures at different
position angles of the beam splitter relative to the sky plane
(for details see \citet{NVF02}). Applying this method, we observed
the magnetic star $\alpha^2$\,CVn which shows rotationally
modulated linear polarization in spectral lines of its spectrum.

\begin{figure}
\centering
\includegraphics[width=8.5cm, height=7.5cm, angle=0]{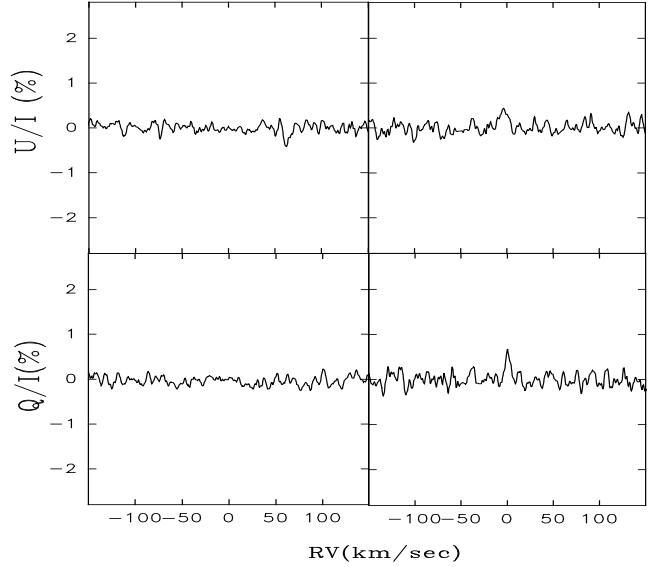}
\vspace*{0.5cm} \caption{Averaged per basic Balmer lines Stokes $Q/U$
parameters of the star
$\alpha^2$\,CVn at those rotational phases where the star
exhibits zero ($\phi \approx 0$, left plots) and maximun
($\phi \approx 0.2$, right plots) linear polarization.
The data are presented in the radial velocity scale as functions
of the distance from the spectral line cores.
} \label{fig11}
\end{figure}

According to \cite{WDL00}, net linear polarization of this star
varies from about -0.2\% to about +0.6\% . We observed
$\alpha^2$\,CVn at two phases of the star's rotation when the
polarization is nearly zero ($\phi \approx 0.03$) and at one of
the extrama ($\phi \approx 0.2$) where the star's spectrum
exhibits non-zero positive linear polarization ($Stokes-Q \approx
+0.6$\%  and $Stokes-U \approx +0.3$\% , see Fig.\,6 in
\citet{WDL00}). Such a weak polarization level can not simply be
registered within consideration of one individual spectral line.
However, the problem can be resolved if several spectral lines are
considered together. Say, considering net polarization as a
function of the distance from the line cores (in the radial
velocity scale), the LSD method of obtaining averaged per all
available spectral lines polarization can be applied
\citep{WDL00}. To illustrate this, in Fig.\,\ref{fig11} we present
averaged per all available Balmer lines $Stokes-Q/U$ spectra of
$\alpha^2$\,CVn. The left two plots illustrate zero polarization
at the rotation phase $\phi \approx 0.03$. The right plots present
nearly maximum positive polarization at the rotation phase $\phi
\approx 0.2$. Signatures of the narrow, about 0.6\% polarization
features are clearly seen at the line cores in this case. As one
can see, even such weak polarization level can also be registered
with our polarimeter that enables us to measure the transverse
magnetic field. This problem, however, deserves additional special
paper where we will present all the necessary tool for these
measurements and details on linear polarization observations of
the standard stars including $\alpha^2$\ CVn that we briefly
touched here. In this paper we limit ourself with presentation of
the polarimeter and consideration of the longitudinal magnetic
field measurements only.

\subsection{Measurements of stellar longitudinal magnetic fields
with the BOES}

In our first Zeeman observations of stellar magnetic fields with
the BOES we measured longitudinal fields via analysis of the
displacement (Equation 1) between the positions of spectral lines
in the spectra of opposite circular polarizations (see, for
instance, \citet{MFV02} and references therein). This technique,
if applied to all selected spectral lines in a spectrum with
further statistical averaging of the result, is quite robust for
the determination of longitudinal magnetic fields averaged over
the stellar disks. The atomic data necessary for the
identification of spectral lines and their Lande factors were
taken from the VALD database \citep{PKR95,RPS99,KPR99}.

Measurements at individual spectral lines were carried out in
pixel coordinate system separately for spectra obtained at the
ordinary and extraordinary beams split by the analyzer to avoid
uncertainties in the wavelength calibration. In this case, Zeeman
displacement between $\sigma$ components of opposite circular
polarizations can be measured as a shift between the centers of
gravity of a spectral line extracted from the same pixels of two
neighboring CCD frames obtained at two orthogonal orientations
($+45^\circ$ and $-45^\circ$) of the quarter-wave plate. The
longitudinal magnetic field $B_l$ based on these measurements can
be found as an averaged mean of $B_l^o$ and $B_l^e$ estimates of
the magnetic field obtained from spectra of the corresponding
ordinary and extraordinary beams. It is clear, that $B_l^o$ and
$B_l^e$ are not free of any biases due to the influence of
mechanical instabilities from an exposure to exposure. Their
averaging, however, significantly reduces them as we now
illustrate.

Denoting the center of gravity of spectral lines from the
ordinary/extraordinary spectrum obtained at $+45^\circ$ or
$-45^\circ$ position of the quarter-wave plate as
$\lambda^o_{+45^\circ}/\lambda^e_{+45^\circ}$ or
$\lambda^o_{-45^\circ} / \lambda^e_{-45^\circ}$, the $B_l^o$ and
$B_l^e$ estimations have the following forms:

\begin{equation}
B_l^o = \pm k(\lambda^o_{+45^\circ} - \lambda^o_{-45^\circ})/2 =
k(\pm 2\Delta \lambda_{B} \pm \Delta \lambda_o)/2
\end{equation}

\begin{equation}
B_l^e = \mp (\lambda^e_{+45^\circ} - \lambda^e_{-45^\circ})/2 =
k(\pm 2\Delta \lambda_{B} \mp \Delta \lambda_e)/2
\end{equation}

\noindent where $k = 1/(4.67\times 10^{-13}z\lambda^2)$; $\Delta
\lambda_B$ is Zeeman displacement between the $\sigma$ components
in the spectral line; $\Delta \lambda_o$ and $\Delta \lambda_e$
are instrumental shifts between spectra of the
ordinary/extraordinary beam obtained at different time moments due
to the requirement to have two consecutive exposures at
$+45^\circ$ and $-45^\circ$ orientations of the quarter-wave
plate. The sign inversion in $B_l^e$ estimation appears due to the
inverse polarimetric properties of the extraordinary beam relative
to the ordinary one. Averaging the data we have:

\begin{equation}
B_l = B_l^o / 2 + B_l^e / 2  = k(\pm 4\Delta \lambda_{B}
\pm \Delta \lambda_o \mp \Delta \lambda_e ) / 4.
\end{equation}

\noindent In eq.(5), the instrumental shift is taken with
opposite sign due to the polarimetric inversion in the beams.
Practically in all cases of any instrumental effect these shifts
are equal in linear guess approximation and eliminated each other
in the applied method.

Obtaining an estimate of the averaged mean longitudinal magnetic
fields and error bars within one observation of a star (consisting
of four consecutive exposures) was done by weighting and
statistically averaging all the individual measurements. Here, we
weighted individual line-by-line measurements by residual
intensities and signal-to-noise ratios as described by
\citet{MFV02}. Alternatively, we also applied the Monte-Carlo
simulation method \citep{pla04a} of obtaining error bars and
weights for the individual measurements.

\section{Results of Zeeman observations of standard stars}

The observations were carried out in the course of one observing
night on 27th September, 2006. Three well known magnetic Ap/Bp
stars HD215441, HD32633 and HD40312 ($\theta Aur$) were observed
together with Procyon as a zero-field standard.
Table~\ref{Tphases} gives an overview of the observations.

Results of the longitudinal magnetic field measurements are
summarized in Table~\ref{Res}, where column~1 is the name of a
star, column~2 is the spectral class, column~3 is the visual
magnitude, column~4 is the exposure time, column~5 is the
rotational phase of a magnetic star if known (throughout this
study we use ephemeris presented by \citet{WDL00}), columns 6-7
report the measurements and uncertainties of the longitudinal
magnetic fields obtained as explained above. We briefly discuss
these results.

{\bf HD215441} or the famous Babcock's star is known to have the
strongest magnetic field among the main sequence stars. The
longitudinal field of this star varies with the rotational period
P~=~9.4871~days about the mean value of  $+15000$\,G with an
amplitude of about 4.5~kG \citep{BBM05}. Our result($B_l = +10500
\pm 330$\,G, see Table~\ref{Res}), which is consistent with the
data presented by \citet{BBM05}, was measured as the average mean
of individual Zeeman measurements obtained from all available 68
spectral lines (including hydrogen lines) in the region between
4100\AA\, and 8000\AA\,. Unfortunately, due to uncertainties in
determination of the magnetic ephemeris of this star \citep{BBM05}
we are unable to compare our result with other authors' ones.
Besides, measurements of $B_l$ by using only metal lines compared
to measurements by hydrogen lines exhibit a very large discrepancy
(much larger than corresponding scatter due to Poisson noise),
that suggests the presence of a very complicated magnetic field
morphology in this star. For example, measurements of the magnetic
field by using only the Balmer lines gives $B_l = +17000 \pm
700$\,G that is confidentially larger than the field averaged by
individual measurements of all the other spectral lines. In this
study we do not discuss this well-known effect \citep[and
references therein]{BBM05} presenting HD215441 only as an
illustration.

{\bf HD32633} and {\bf HD40312} or {\bf $\theta Aur$} are
well-known broad-lined B9p and A0p magnetic stars which we used as
standards to compare our measurements with the best measurements
given by other authors. To our knowledge, the most extensive
high-precision measurements of their variable magnetic fields were
performed by \citet{WDL00}.

\begin{figure}
\centering
\includegraphics[width=8.5cm, height=8.5cm, angle=0]{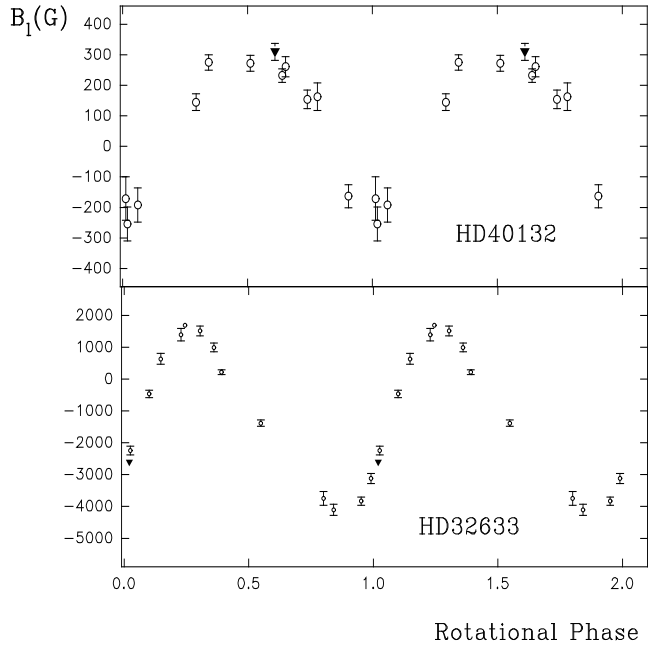}
\vspace*{0.5cm} \caption{ Comparison of our Zeeman observations of
longitudinal magnetic fields of the stars HD40312 and HD32633
(filled triangles) with observations of \citet{WDL00} (open
circles). } \label{fig12}
\end{figure}

According to \citet{WDL00}, the longitudinal magnetic field of
HD32633 varies with the period of 6.4300~days and demonstrates
smooth, non-sinusoidal variation with two extremes at $B_l \approx
-4200$\,G and $B_l \approx +1800$\,G. Our single observation of
this star was carried out at the rotation phase $\phi = 0.012$
under the moderate weather conditions. At this phase the
longitudinal magnetic field of this star is located at arising
branch of the field variation between the negative extremum and
crossover. With our polarimeter, the measured longitudinal field
at this phase ($B_l = -2616 \pm 56$\,G, see Table~\ref{Res}) have
demonstrated a very good agreement with results of other authors.
In Fig.\,\ref{fig12} (the lower plot) our observation (filled
triangles) is illustrated in comparison with the data (open
circles) taken from \citet{WDL00}. The obtained field value and
small error bar (in Fig.\,\ref{fig12} the error bar is smaller
than the triangle size) suggest high quality of our polarimeter in
measurements of stellar magnetic fields. In particular, the
previous best estimations of the mean longitudinal magnetic field
of this star demonstrated the same or twice lower accuracy
\citep{WDL00}.

The star {\bf HD40312} exhibits weak variable longitudinal
magnetic field the behavior of which is presently well-studied
\citep{WDL00}. Our observations of {\bf HD40312} were carried out
under poor weather condition that made us observe this star for a
quite long time (about one hour, Table~\ref{Tphases}). We have
observed this target at the rotation phase $\phi \approx 0.6$ of
the field variation. Similar to the previous star, our estimation
of the field is well consistent with results given by
\citet{WDL00} (see Table~\ref{Res} and the upper plot in
Fig.\,\ref{fig12}).

{\bf HD61421} (or {\bf $\alpha$\,CMi}, {\bf Procyon}) was finally
used as a zero-field star. This star was observed polarimetrically
by a number of authors (see an overview in Table~\ref{Alp}). In
their observations magnetic field has not been found with typical
accuracies from about 1~G to 7~G. Formal averaging of these
results gives no any traces of magnetic field at a level of about
0.5~G that makes us possible to use this star as a well-studied
zero-field standard. Our result also showed a very accurate zero
with error bar of about 2\,G (see Table~\ref{Res}). This result
has been obtained during just about 6~min of integration in
contrast to hours of typical exposure times in the previous
observations of this star (see Table~\ref{Alp}). About one hundred
non-blended spectral lines were used to obtain this result.

It should be noted here that such a high accuracy of the field
measurement becomes critical to the adopted methods of data
reduction and the measurements. In our measurements, for example,
different methods of statistical analysis in the applied
line-by-line technique gave us some differences in the final
result. Obtaining statistical weights of the measurements at
individual spectral lines using their residual intensities and
signal-to-noise ratios, we initially derived the uncertainty of
about 2.7\,G. However, from the measurements weighted by
Monte-Carlo simulation method \citep{pla04a}, the final error bar
was determined with significantly better accuracy of about 2.2\,G.
This fact can easily be understood because there are more than the
above mentioned two factors (poisson noise and residual
intensities) influence to the statistical weights. The Lande
factors (their values) and shapes of spectral lines also play
roles. The applied Monte-Carlo method takes them all into account.

At the same time, methods of weighting do not play such a
significant role in the determination of strong mean longitudinal
fields in chemically peculiar magnetic stars. In these stars the
main contribution to the uncertainty comes from the inhomogeneous
distribution of the magnetic field over the surface and chemical
peculiarities that may demonstrate local field intensities at
chemically overabundant spots (such as in our example with
HD215441).

Different methods of the data reduction also play a role
(extraction of spectral orders, for example). In this paper,
however, traditional standard methods were adopted for the
spectropolarimetric analysis to demonstrate the pilot workability
of the BOES spectropolarimeter.

\section{Summary}

We have presented the new stationary spectropolarimeter mounted on
the high-resolution fiber-fed echelle spectrograph BOES of the
1.8-m telescope at the BOAO. At the moment the instrument is ready
for regular observations of stellar longitudinal magnetic fields
and demonstrates good transparency and precision of the
measurements. Typical accuracies of the field measurements in
bright stars (brighter than 9th stellar magnitude) range from
about 2 to a few tens of gauss depending on spectral class,
rotation of a star and integration time. In this part, further
improvements to the polarimeter concern mainly the software for
the data reduction and methods of the measurements. We expect,
that applying more advanced technologies of the data reduction
connected, for example, with optimal extraction of spectral orders
\citep{DSC97} or the LSD method for measurements of weak magnetic
fields \citep{DSC97}, the instrument will demonstrate even better
results. Further examination of these points will be among the
goals of our final study upon final upgrade of the polarimeter.

\acknowledgments ACKNOWLEDGMENTS

This work was supported by Korea Astronomy and Space Science
Institute (KASI) under grant No. 2007-1-310-A0.  We wish to thank
our anonymous referee for comments on the first draft of this
paper, which lead to considerable improvement of this paper.  GV
is grateful to the Korean MOST (Ministry of Science and
Technology, grant M1-022-00-0005) and KOFST (Korean Federation of
Science and Technology Societies) for providing him an opportunity
to work at KASI through Brain Pool program. And BCL acknowledges
his work as part of the research activity of the Astrophysical
Research Center for the Structure and Evolution of the Cosmos
(ARCSEC, Sejong University) of the Korea Science and Engineering
Foundation (KOSEF) through the Science Research Center (SRC)
program.

\newpage

\begin{table}
\caption{Journal of observations. The first column gives name of
an observed star, the second column lists the Julian date {\bf
(JD)} of the midpoint of the observation, the third and fourth
columns are total exposure time {\bf Exp} and corresponding number
of observations {\bf N}$_{obs}$, fifth and sixth columns are total
signal to noise ratio {\bf S/N} at $\lambda 5500 $ and weather
conditions ({\bf WC}: good/moderate/poor).} \vspace*{0.5cm}
\centering
\begin{tabular}{cccccc}
\hline\hline
Name & JD & Exp (sec) & N$_{obs}$ & S/N & WC\\
\hline
 HD32633  & 2454006.065278  & 8400 & 1 & 350 & moderate\\
 HD40312  & 2454006.256944  & 3840 & 3 & 400 & poor\\
 HD61421 & 2454006.333333  & 395  & 1 & 600 & poor\\
 HD215441 & 2454005.981944  & 7200 & 1 & 180 & moderate\\
\hline
\end{tabular}
\label{Tphases}
\end{table}

\begin{table}
\caption{Determination of the mean longitudinal field.
}
\vspace*{0.5cm}
\centering
\begin{tabular}{ccccccc}
\hline\hline
Name & SP & m$_{v}$ & Exp (sec) & $\phi$ & B$_l$ (G)& $\sigma$ (G)\\
\hline
 HD32633  & B9p  & 7.1 & 8400 & 0.13 & -2616  & 56\\
 HD40312  & A0p  & 2.6  & 3840 & 0.61  & +310   & 28\\
 HD61421 & F5    & 0.34  & 395  & --   & -3.8   & 2.2\\
 HD215441 & B9p  & 8.8  & 7200 & --   & +10500 & 330\\
\hline
\end{tabular}
\label{Res}
\end{table}

\begin{table}
\caption{An overview of the mean longitudinal magnetic field observations of $\alpha$ CMi.}
\vspace*{0.5cm}
\centering
\begin{tabular}{lclccc}
\hline\hline
{Author}&{Aperture}&{Polarimeter}&{$Exposure$}&{$B_{e}$}&{$\sigma$}\\
{}&{(meters)}&{}&{(hours)}&{(G)}&{(G)}\\
\hline
\cite{land82}   & 2.6 & magnetometer               & ?   & 7     &  7 \\
\cite{bor84}    & 2.5 & multislit magnetometer     & ?   & -7.5  &  5.9\\
\cite{gl91}     & 6.0 & magnetometer               & ?   & 17    &  7.1\\
\cite{bed95}    & 1.9 & triple magnetooptical      & 8   & -1.86 &  0.9\\
                &     & filter with a potassium cell&    & 0.49  &  0.8\\
\cite{pla99}    & 2.6 & stokesmeter + CCD          & 2.2 & -1.34 &  1.0\\
\cite{sho02}    & 2.0 & stokesmeter + CCD          & ?   & 2.0   &  5.0\\
\hline
\end{tabular}
\label{Alp}
\end{table}

\end{document}